\documentclass[aps,prd,twocolumn,showpacs,amsmath,amssymb]{revtex4-1}

\usepackage[dvips]{graphicx}
\usepackage[retainorgcmds]{IEEEtrantools}

\begin{document}

\title
{
  Fake evolution of dark energy from observation data
}
\author{Shi Qi}
\email{qishi11@gmail.com}
\affiliation
{
  Purple Mountain Observatory,
  Chinese Academy of Sciences,
  Nanjing 210008,
  China
}
\affiliation
{
  Key Laboratory of Dark Matter and Space Astronomy,
  Chinese Academy of Sciences,
  Nanjing 210008,
  China
}
\affiliation
{
  Joint Center for Particle, Nuclear Physics and Cosmology,
  Nanjing University---Purple Mountain Observatory,
  Nanjing 210093,
  China
}
\affiliation
{
  Kavli Institute for Theoretical Physics China,
  Chinese Academy of Sciences,
  Beijing 100190,
  China
}

% \date{\today}

\begin{abstract}
  The equation of state (EOS) of the dark energy is the key parameter
  to study the nature of the dark energy from the observation.
  Though the dark energy is found to be well consistent with the
  cosmological constant with a constant EOS of $-1$,
  weak evidences from different observation data and analyses show
  that dark energy models with an evolving EOS slightly less than $-1$
  at some medium redshifts and greater than $-1$ at high redshifts are
  more favored.
  In this paper, It is shown that how such a pattern of an evolving
  dark energy EOS can be just biases arising from the statistical
  method widely adopted in data analyses together with the dependence
  of the cosmic expansion on the dark energy EOS.
  The issue is actually not limited to dark energy or cosmology.
  It represents a class of mathematical problems of Bayesian
  analysis.
  It should be paid attention to in similar data analyses to avoid
  biases in drawing conclusions.
\end{abstract}


% \keywords{}

\maketitle

\section{Introduction}

Since the discovery the accelerating expansion of the
universe~\cite{Riess:1998cb,Perlmutter:1998np},
which is attributed to the dark energy,
the equation of state (EOS) of the dark energy have been, and will be
for a long time in the future, the key parameter to study the nature
of the dark energy from the observation.
Though a constant dark energy EOS of $-1$ which corresponds to the
cosmological constant is well consistent with the observation
(see e.g.~\cite{Ade:2015xua, Ade:2015rim} for recent analyses),
weak evidences are persistently observed from different data sets
that the dark energy EOS is slightly less than $-1$ at medium
redshifts and greater than $-1$ at high redshifts.
These evidences come from independent analyses of different kinds of
data sets, including both standard candles
like type Ia supernovae (SNe Ia) and gamma-ray bursts (GRBs)
and standard rulers like baryon acoustic oscillations (BAO).
See~\cite{Zhao:2012aw} for a comprehensive study showing this trend
with SNe Ia, BAO, and other data sets
and~\cite{Qi:2008zk, Qi:2009yr, Wang:2011bx} for a series of
studies showing that GRBs favor a dark energy EOS greater than $-1$ at
high redshifts.
See also~\cite{Hu:2014ega, Zheng:2014ara} for some recent studies with
similar results.
Interestingly, even mock data generated by assuming a $\Lambda$CDM
cosmological model also show weak evidences for such a
trend~\cite{Sullivan:2007tx}.
In this paper, It is shown that how the trend can be just biases
resulted from the statistical method widely adopted in data analyses
together with the dependence of the cosmic expansion on the dark
energy EOS.

\section{Luminosity distance and standard candles}

We start from constructing the procedure for estimating constraints of
standard candles on cosmological parameters using mock data.
Consider a luminosity relation of the following form:
\begin{equation}
  \label{eq:lr}
  y = c_0 + \sum_i c_i x_i + \varepsilon
  ,
\end{equation}
where $x_i$s are some luminosity indicators which can be directly
measured from observation,
$\varepsilon$ is a random variable accounting for the intrinsic
scatter $\sigma_{\mathrm{int}}$ of the relation,
and $y$ has the form of
\begin{equation}
  \label{eq:y_dl}
  y = \log
  \left(
    4 \pi d_L^2 \mathcal{F}
  \right)
  ,
\end{equation}
where $d_L$ is the luminosity distance and $\mathcal{F}$ may be any
physical quantity that can be directly measured from observation.
The luminosity relation of Eq.~(\ref{eq:lr}) has incorporated all the
GRB correlations summarized in~\cite{Qi:2011tr}, as well as the
relations used to derived distance moduli from SNe Ia.
For example, the SALT2 method~\cite{Guy:2007dv} gives
$\mu_B = m_B^{\ast} - M + \alpha \cdot x_1 - \beta \cdot c$.
% \begin{equation}
%   % \label{eq:abc}
%   \mu_B = m_B^{\ast} - M + \alpha \cdot x_1 - \beta \cdot c
%   .
% \end{equation}
Let $y = (\mu_B - m_B^{\ast}) / 2.5$, it can be rewritten in the form
of Eq.~(\ref{eq:lr}).
Note that the intrinsic scatter in distance modulus should be divided
by $2.5$ to convert to the intrinsic scatter $\sigma_{\mathrm{int}}$
in $y$.

From a similar derivation like that in~\cite{Agostini:2005fe}, we
know that, for a given sample of standard candles,
the joint likelihood function for the coefficients $c$, the intrinsic
scatter $\sigma_{\mathrm{int}}$, and the cosmological parameters
$\theta$ is
\begin{equation}
  % \label{eq:abc}
  \mathcal{L}(c, \sigma_{\mathrm{int}}, \theta)
  = k \prod_i \mathcal{L}_i
  ,
\end{equation}
where $k$ is the normalization factor, $i$ runs over the standard
candles, and
\begin{IEEEeqnarray}{rCl}
  \label{eq:L_i}
  \mathcal{L}_i
  &=&
  \frac{1}
  {
    \sqrt
    {
      \sigma_{\mathrm{int}}^2 + \sigma_{y_i}^2
      + \sum_j c_j^2 \sigma_{x_{j, i}}^2
    }
  }
  \nonumber
  \\
  & &
  \times
  \exp
  \left[
    -
    \frac
    {
      \left(
        y_i - c_0 - \sum_j c_j x_{j, i}
      \right)^2
    }
    {
      2
      \left(
        \sigma_{\mathrm{int}}^2 + \sigma_{y_i}^2
        + \sum_j c_j^2 \sigma_{x_{j, i}}^2
      \right)
    }
  \right]
\end{IEEEeqnarray}
with $j$ running over the luminosity indicators in
Eq.~(\ref{eq:lr}).
This likelihood function is a quite general one with the well known
$\chi^2$ statistic being a special case of it.

In the discussions below, we ignore the measurement uncertainties
($\sigma_{x_{j, i}}$ and $\sigma_{y_i}$ in Eq.~(\ref{eq:L_i})) to
simplify the problem.

First consider the simplest case. For the luminosity relation of
\begin{equation}
  % \label{eq:abc}
  y = a + b x + \varepsilon
  ,
\end{equation}
to estimate its constraints on the cosmological parameters
$\theta$ using mock data, we follow the following steps:
\begin{enumerate}
\item Set the fiducial values for the parameters of the luminosity
  relation $a$, $b$, $\sigma_{\mathrm{int}}$ and the cosmological
  parameters $\theta$.
  Here, we use $a_0$, $b_0$, $\sigma_{\mathrm{int}, 0}$, and
  $\theta_0$ to denote the fiducial values for the corresponding
  parameters.
\item Generate the mock data for a sample of standard candles.
  Assume the total number of standard candles is $N$.
  For the $i$th standard candles, we generate its redshift $z_i$ and
  luminosity indicator $x_i$ and draw a sample $\varepsilon_i$ from
  the distribution of random variable $\varepsilon$, i.e., the normal
  distribution $\mathcal{N}(0, \sigma_{\mathrm{int}, 0}^2)$.
  From them we know its fiducial value for $y$:
  \begin{equation*}
    y_{i, 0} = a_0 + b_0 x_i + \varepsilon_i
    .
  \end{equation*}
  Then we calculate $\mathcal{F}_i$, i.e. the value of $\mathcal{F}$
  for individual standard candles that should be measured from the
  observation given the above information, through
  \begin{equation*}
    y_{i, 0} = \log
    \left[
      4 \pi d_L^2(z_i, \theta_0) \mathcal{F}_i
    \right]
    .
  \end{equation*}
  Thus, we can calculate $y_i$ for any given cosmological parameters:
  \begin{IEEEeqnarray*}{rCl}
    y_i &=& \log
    \left[
      4 \pi d_L^2(z_i, \theta) \mathcal{F}_i
    \right]
    \\
    &=& \log
    \left[
      4 \pi d_L^2(z_i, \theta_0) \mathcal{F}_i
    \right]
    \\
    & &
    - \log
    \left[
      d_L^2(z_i, \theta_0)
    \right]
    + \log
    \left[
      d_L^2(z_i, \theta)
    \right]
    \\
    &=&
    y_{i, 0}
    + 2 \log
    \left[
      d_L(z_i, \theta) / d_L(z_i, \theta_0)
    \right]
    \\
    &=&
    a_0 + b_0 x_i + \varepsilon_i
    + 2 \log
    \frac
    {
      d_L(z_i, \theta)
    }
    {
      d_L(z_i, \theta_0)
    }
    .
  \end{IEEEeqnarray*}
  For later convenience, we define
  \begin{equation}
    \label{eq:lztheta_dl}
    l(z, \theta, \theta_0)
    =
    2 \log
    \frac
    {
      d_L(z, \theta)
    }
    {
      d_L(z, \theta_0)
    }
  \end{equation}
  and use $l_i$ as the abbreviation for $l(z_i, \theta, \theta_0)$.
  So, we have
  \begin{equation*}
    y_i = a_0 + b_0 x_i + \varepsilon_i + l_i
    .
  \end{equation*}
\item Calculate the likelihood.
  Ignoring the measurement uncertainties, we have
  \begin{IEEEeqnarray*}{rCl}
    \mathcal{L}_i
    &=&
    \frac{1}{\sigma_{\mathrm{int}}}
    \exp
    \left[
      - \frac
      {
        (y_i - a - b x_i)^2
      }
      {
        2 \sigma_{\mathrm{int}}^2
      }
    \right]
    \\
    &=&
    \frac{1}{\sigma_{\mathrm{int}}}
    \exp
    \left[
      - \frac
      {
        \left(
          \delta a + \delta b x_i
          - \varepsilon_i - l_i
        \right)^2
      }
      {
        2 \sigma_{\mathrm{int}}^2
      }
    \right]
    ,
  \end{IEEEeqnarray*}
  where $\delta a \equiv a - a_0$ and $\delta b \equiv b - b_0$.
  Let $m_i = \delta b x_i - \varepsilon_i - l_i$, the joint likelihood
  function is
  \begin{IEEEeqnarray*}{rCl}
    \IEEEeqnarraymulticol{3}{l}
    {
      \mathcal{L}(a, b, \sigma_{\mathrm{int}}, \theta)
      =
      k \prod_{i=1}^N \mathcal{L}_i
    }
    \\
    &=&
    \frac{k}{\sigma_{\mathrm{int}}^N}
    \exp
    \left[
      - \frac
      {
        \sum_{i=1}^N
        \left(
          \delta a + m_i
        \right)^2
      }
      {
        2 \sigma_{\mathrm{int}}^2
      }
    \right]
    \\
    &=&
    \frac{k}{\sigma_{\mathrm{int}}^N}
    \exp
    \left[
      - \frac
      {
        (\delta a)^2
        + 2 \overline{m} \delta a
        + \overline{m^2}
      }
      {
        2 \sigma_{\mathrm{int}}^2 / N
      }
    \right]
    ,
  \end{IEEEeqnarray*}
  where $\overline{m}$ and $\overline{m^2}$ are the average values of
  $m$ and $m^2$ over the standard candles.
\item Marginalize out the nuisance parameters.
  Since we aim at constraining the cosmological parameters $\theta$,
  we need integrate the joint likelihood function over the nuisance
  parameters.
  In this case, they are the calibration parameters
  $a$, $b$, and $\sigma_{\mathrm{int}}$.
  \begin{enumerate}
  \item \label{enum:intercept} Integrate over the intercept
    parameter.
    \begin{IEEEeqnarray*}{rCl}
      \IEEEeqnarraymulticol{3}{l}
      {
        \mathcal{L}(b, \sigma_{\mathrm{int}}, \theta)
        =
        \int_{-\infty}^{+\infty}
        \mathcal{L}(a, b, \sigma_{\mathrm{int}}, \theta)
        \mathrm{d} \delta a
      }
      \\
      &=&
      \frac{k}{\sigma_{\mathrm{int}}^{N-1}}
      \sqrt{\frac{2 \pi}{N}}
      \exp
      \left(
        - \frac
        {
          \sigma_m^2
        }
        {
          2 \sigma_{\mathrm{int}}^2 / N
        }
      \right)
      ,
    \end{IEEEeqnarray*}
    where
    $\sigma_m^2
    = \frac{1}{N} \sum_{i=1}^N (m_i - \overline{m})^2
    = \overline{m^2} - \overline{m}^2$
    is the variance of $m$.
  \item \label{enum:slope} Integrate over the slope parameter.
    Let $n_i = \varepsilon_i + l_i$, then
    $m_i = \delta b x_i - n_i$.
    Since $x_i$, $\varepsilon_i$, and $l_i$ are independent of
    each other, we have
    \begin{equation*}
      \sigma_m^2 = (\delta b)^2 \sigma_x^2 + \sigma_n^2
      .
    \end{equation*}
    Therefore
    \begin{IEEEeqnarray*}{rCl}
      \IEEEeqnarraymulticol{3}{l}
      {
        \mathcal{L}(\sigma_{\mathrm{int}}, \theta)
        =
        \int_{-\infty}^{+\infty}
        \mathcal{L}(b, \sigma_{\mathrm{int}}, \theta)
        \mathrm{d} \delta b
      }
      \\
      &=&
      \frac{k}{\sigma_{\mathrm{int}}^{N-2}}
      \frac{2 \pi}{N}
      \frac{1}{\sigma_x}
      \exp
      \left(
        - \frac
        {
          \sigma_n^2
        }
        {
          2 \sigma_{\mathrm{int}}^2 / N
        }
      \right)
      .
    \end{IEEEeqnarray*}
    Note that $\sigma_x$ here is the standard deviation of $\{x_i\}$,
    not the measurement uncertainty $\sigma_{x_{j, i}}$ in
    Eq.(\ref{eq:L_i}).
  \item Integrate over the intrinsic scatter.
    \begin{IEEEeqnarray*}{rCl}
      \IEEEeqnarraymulticol{3}{l}
      {
        \mathcal{L}(\theta)
        =
        \int_{0}^{+\infty}
        \mathcal{L}(\sigma_{\mathrm{int}}, \theta)
        \mathrm{d} \sigma_{\mathrm{int}}
      }
      \\
      &=&
      \frac{k \pi}{N \sigma_x}
      2^{\frac{N-3}{2}}
      \Gamma \left( \frac{N-3}{2} \right)
      \left(
        N \sigma_n^2
      \right)^{- \frac{N-3}{2}}
      \\
      &\propto&
      \left(
        \sigma_n^2
      \right)^{- \frac{N-3}{2}}
      .
    \end{IEEEeqnarray*}
    Here $\sigma_n^2 = \sigma_{\varepsilon}^2 + \sigma_l^2$, and
    $\sigma_{\varepsilon}^2$ is an estimation of
    $\sigma_{\mathrm{int}, 0}^2$.
    So the marginal likelihood of the cosmological parameters $\theta$
    is given by
    \begin{equation*}
      \mathcal{L}(\theta)
      \propto
      \left(
        \sigma_{\mathrm{int}, 0}^2 + \sigma_l^2
      \right)^{- \frac{N-3}{2}}
      .
    \end{equation*}
  \end{enumerate}
\end{enumerate}

For the general case of Eq.~(\ref{eq:lr}), the steps and the
derivation are basically the same as the above.
The step~\ref{enum:slope} can be repeated until all the slope
parameters are integrated over.
It is easy to check that the generalized marginal likelihood of the
cosmological parameters $\theta$ is given by
\begin{equation}
  \label{eq:Ltheta}
  \mathcal{L}(\theta)
  \propto
  \left(
    \sigma_{\mathrm{int}, 0}^2 + \sigma_l^2
  \right)^{- \frac{N-p}{2}}
  ,
\end{equation}
where $p$ is the number of the calibration parameters which include
the coefficients $c$ and the intrinsic scatter
$\sigma_{\mathrm{int}}$.
This also applies to the case of no luminosity indicator, i.e., the
case of the luminosity relation $y = a + \varepsilon$.

Thus, at this point, given a luminosity relation, we do not need to
follow the above steps any more to estimate its constraints on
cosmological parameters.
Instead, we can directly calculate the marginal likelihood of the
cosmological parameters using
Eqs.~(\ref{eq:lztheta_dl}) and (\ref{eq:Ltheta})
with the input of the number of luminosity indicators involved, the
intrinsic scatter of the luminosity relation, and the number of the
standard candles and their redshifts.

Throughout this paper to the end, if not stated explicitly otherwise,
the flat $\Lambda$CDM with $\Omega_m = 0.3$ is used as the fiducial
cosmological model.
And since our focus is on the dark energy EOS $w(z)$, we fix all other
cosmological parameters at their fiducial values.
Thus $l(z, \theta, \theta_0)$ reduces to $l(z, w(z), -1)$.
For the $w$CDM cosmological model where $w(z)=w$ is a constant
along the redshift, it further reduces to $l(z, w, -1)$.

\begin{figure}[htbp]
  \centering
  \includegraphics[width = 0.23 \textwidth]{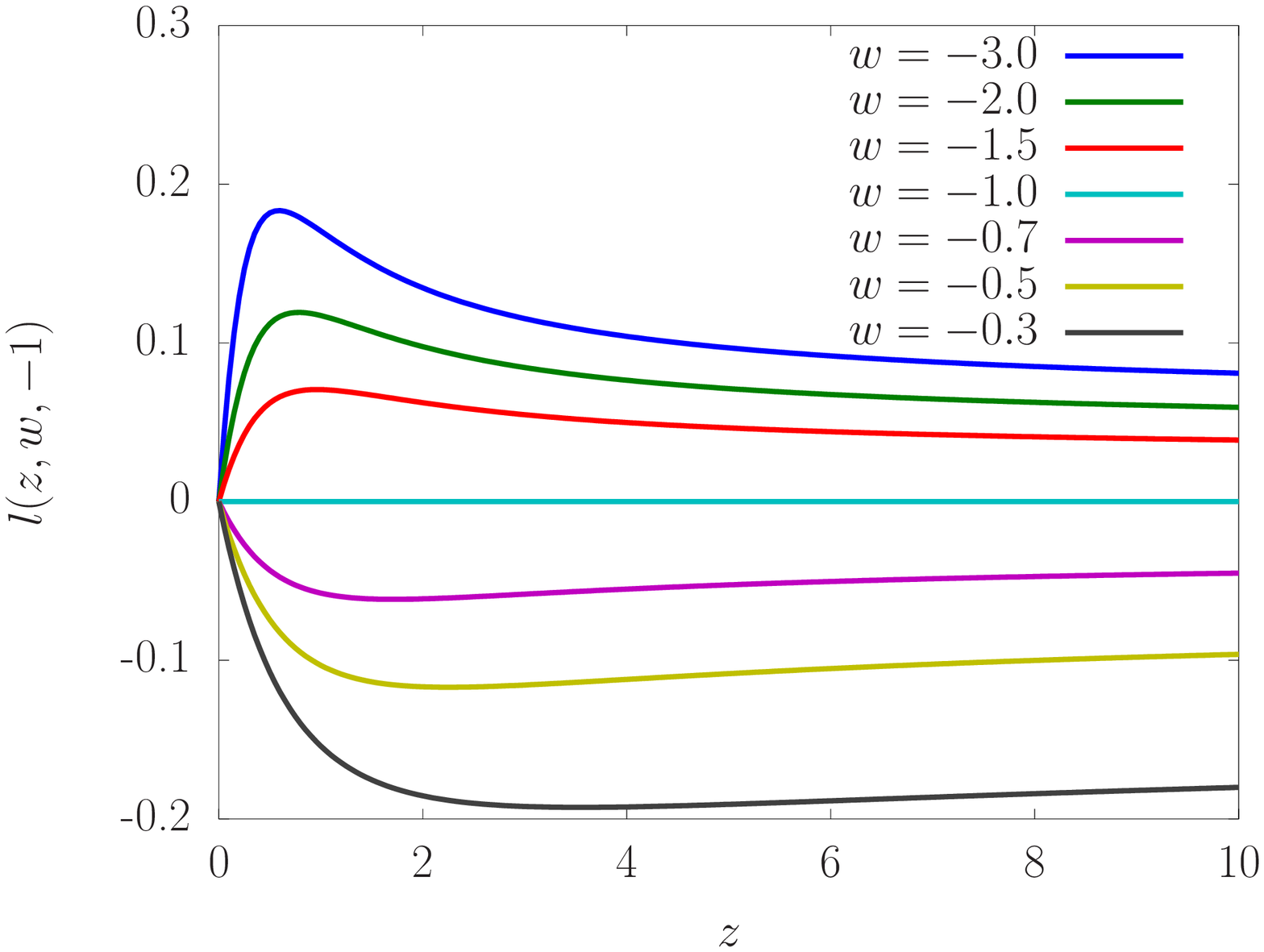}
  \includegraphics[width = 0.23 \textwidth]{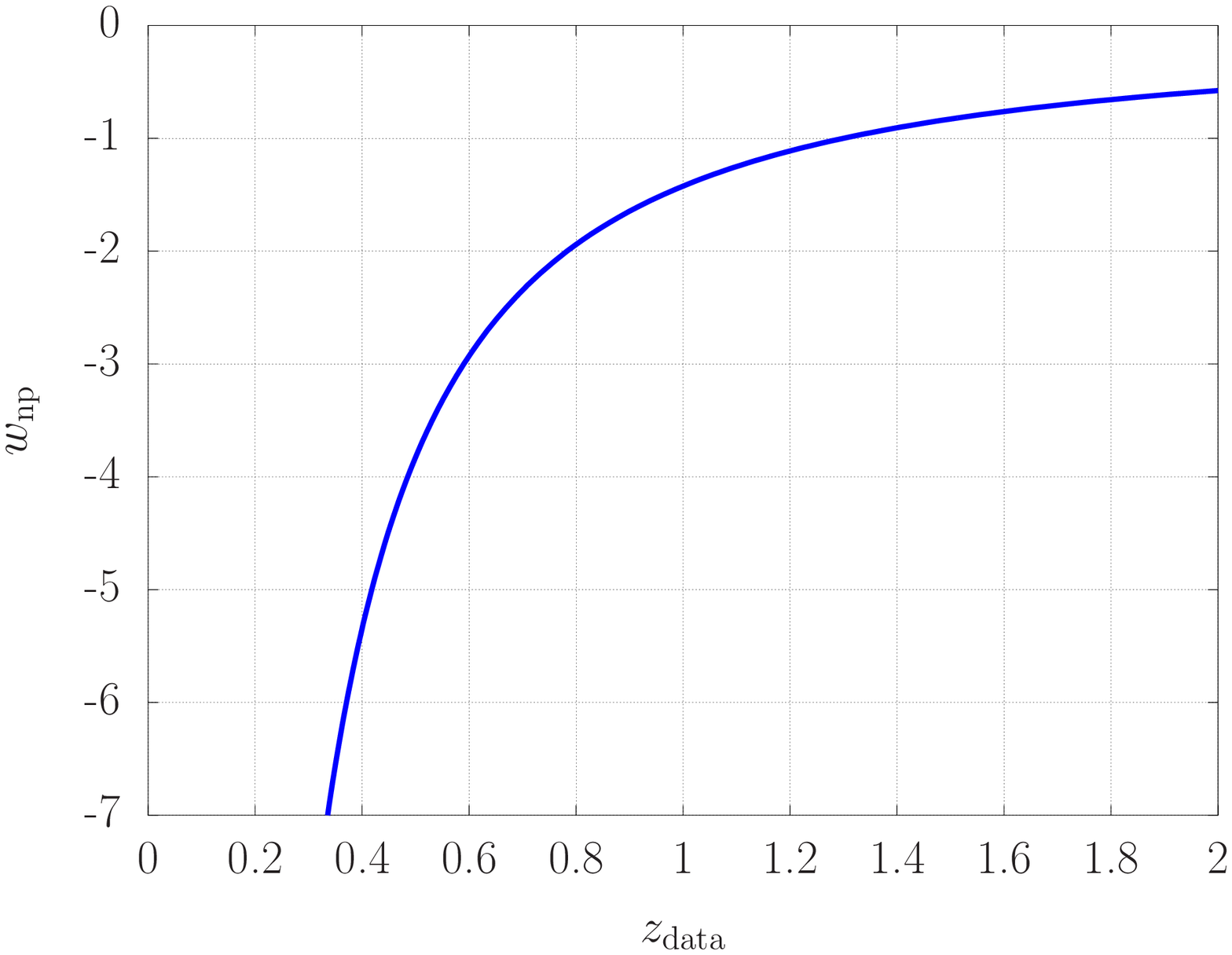}
  \caption
  {
    $l(z, w, -1)$ versus $z$ for some values of $w$
    and
    $w_{\mathrm{np}}$ versus $z_{\mathrm{data}}$
    for standard candles.
  }
  \label{fig:lw_sc}
\end{figure}

\begin{figure}[htbp]
  \centering
  \includegraphics[width = 0.23 \textwidth]{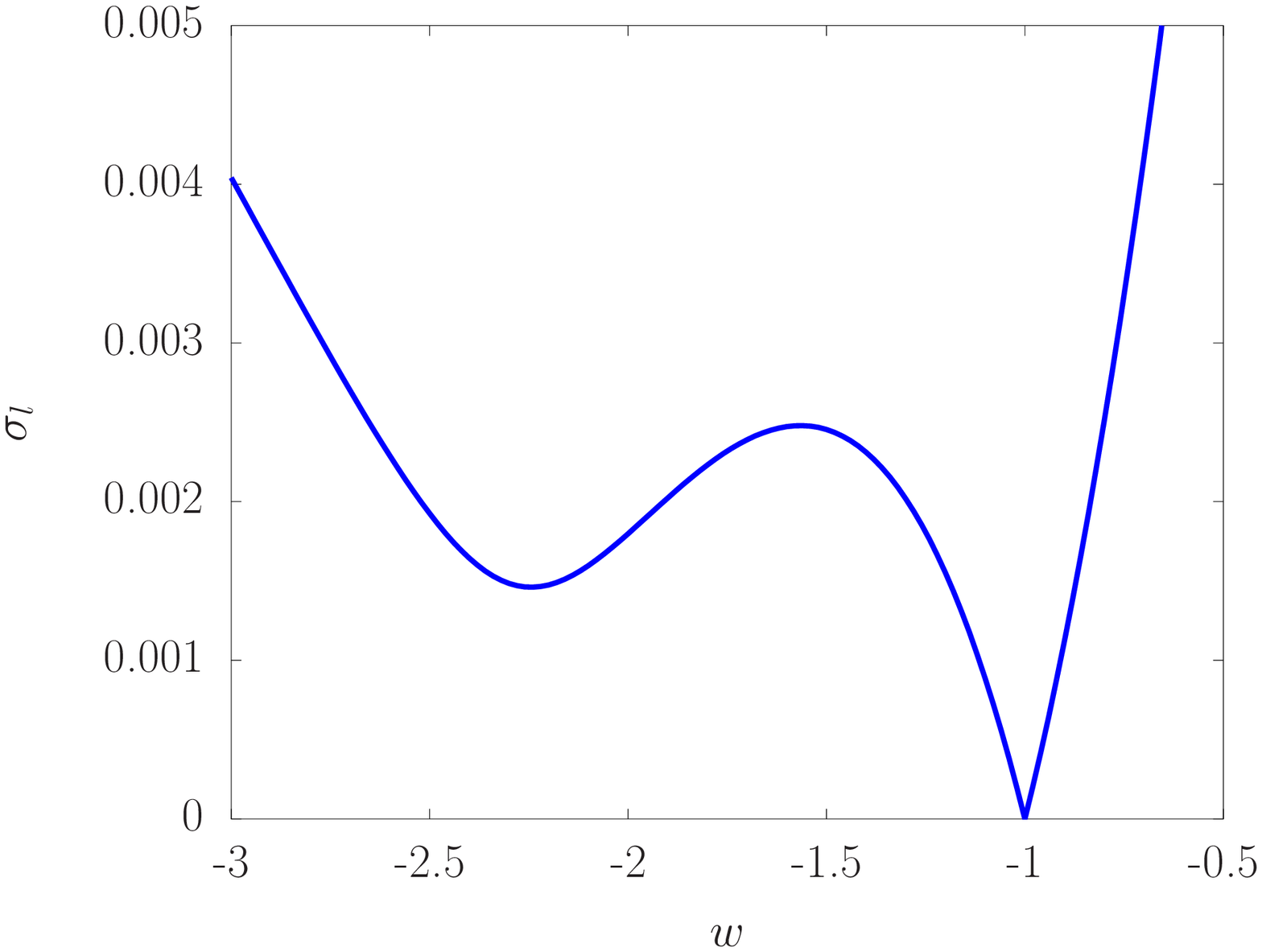}
  \includegraphics[width = 0.23 \textwidth]{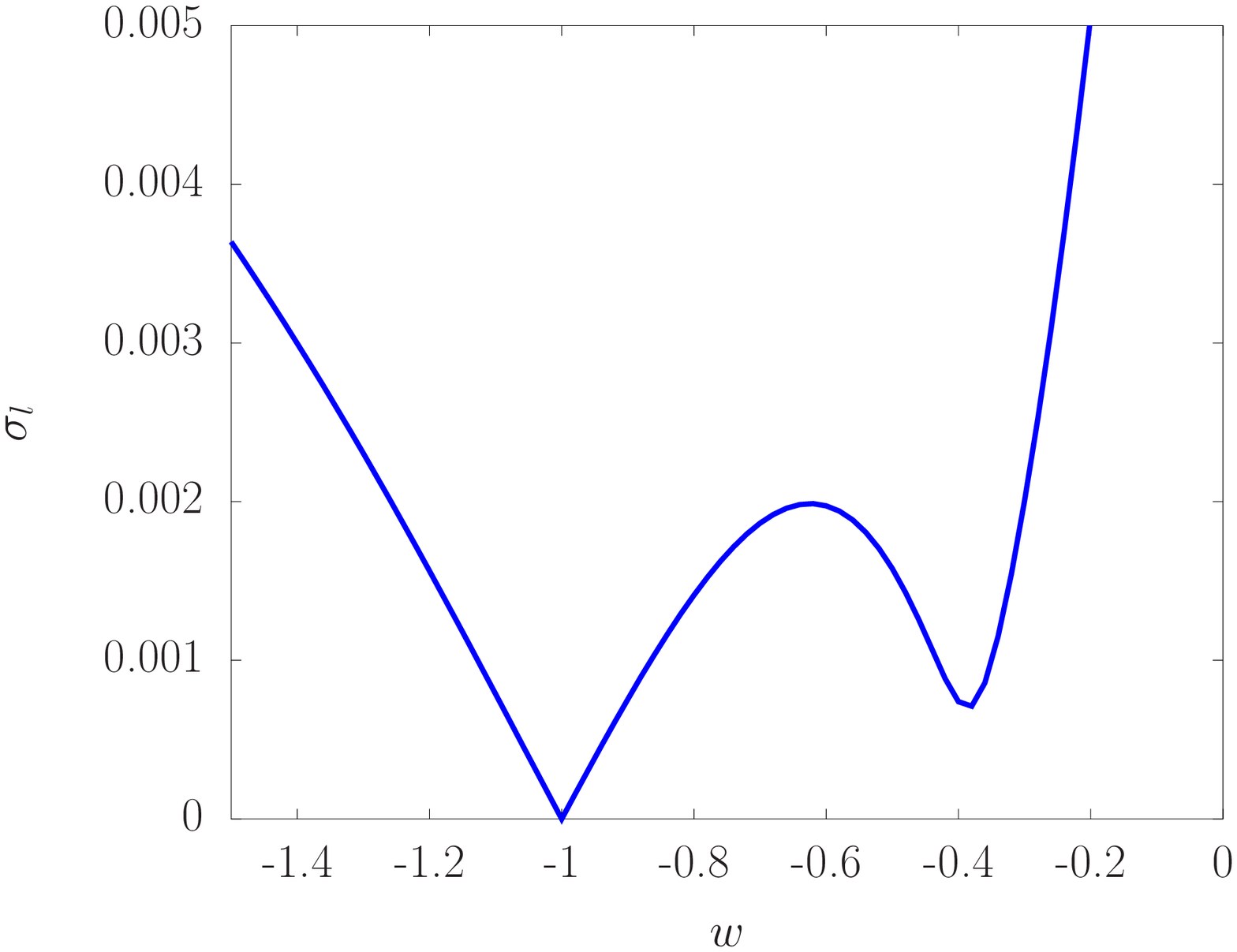}
  \\
  \includegraphics[width = 0.23 \textwidth]{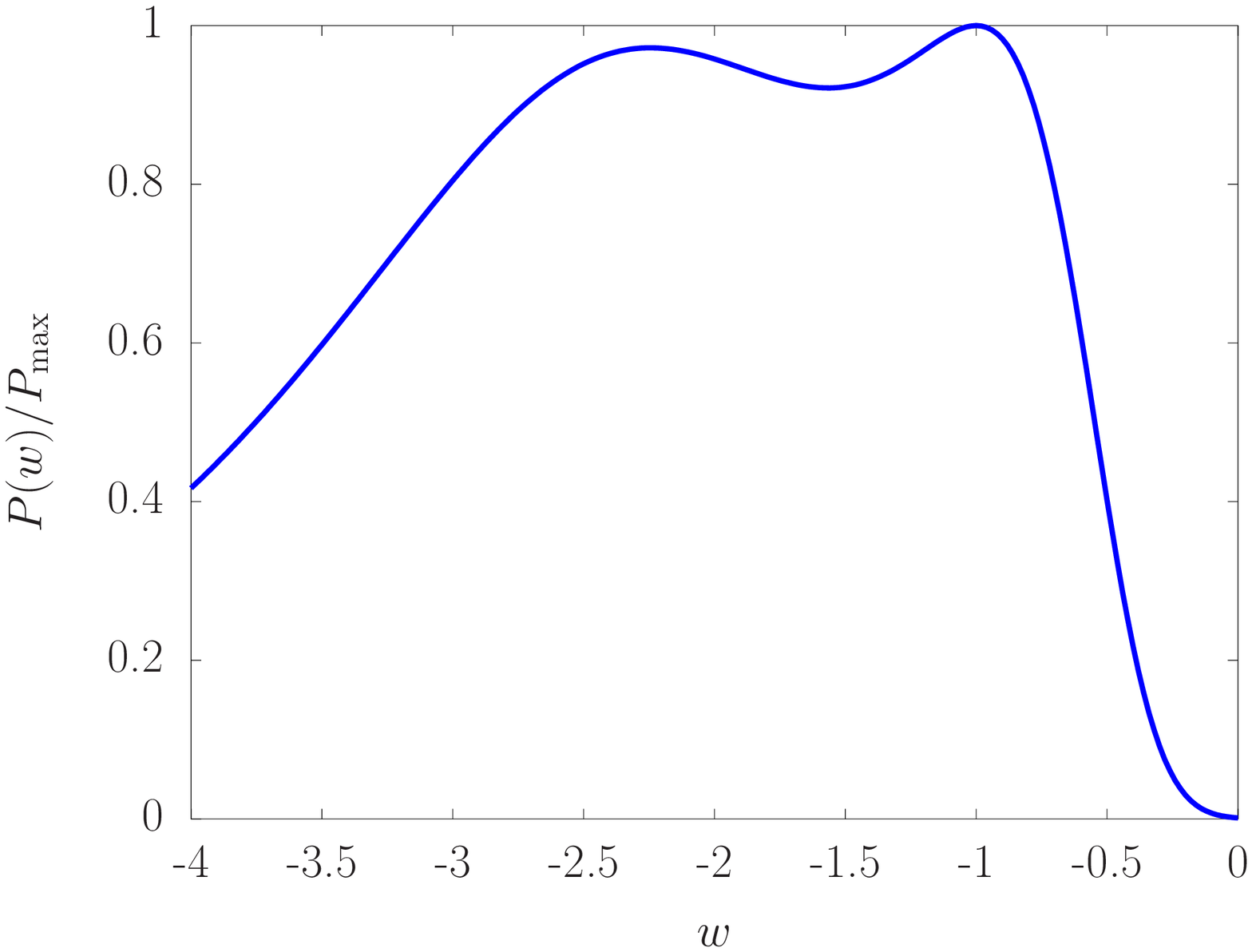}
  \includegraphics[width = 0.23 \textwidth]{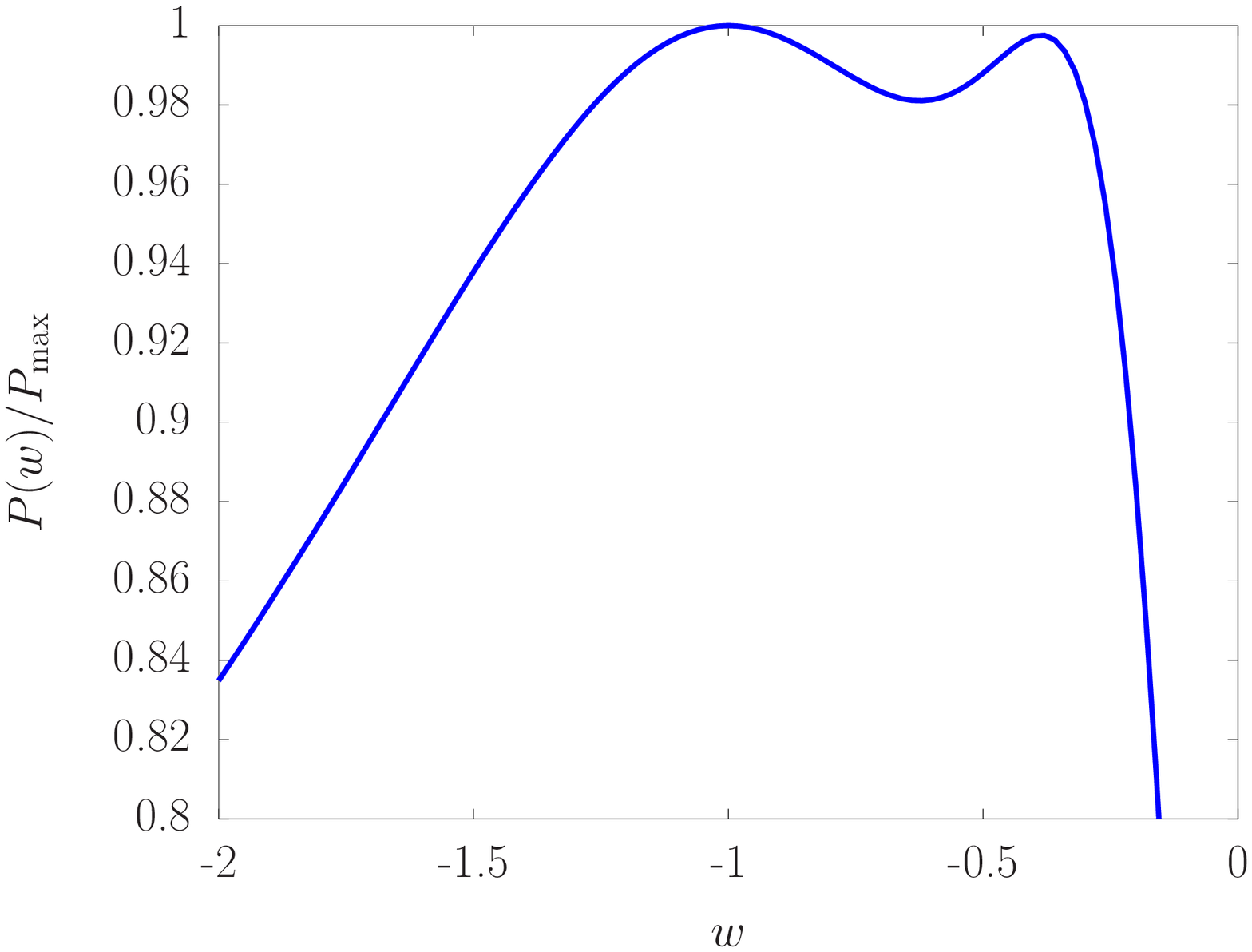}
  \caption
  {
    $\sigma_l$ versus $w$ and the corresponding probability
    distribution of $w$.
    For the left column, $100$ standard candles uniformly distributing
    in the redshift range $[0.5, 1]$ are used,
    and $\sigma_{\mathrm{int}, 0} = 0.06$ and $p = 4$ are assumed for
    the luminosity relation.
    For the right column, $100$ standard candles uniformly
    distributing in the redshift range $[2, 4]$ are used,
    and $\sigma_{\mathrm{int}, 0} = 0.1$ and $p = 3$ are assumed for
    the luminosity relation.
  }
  \label{fig:sl_P_w}
\end{figure}

Now let us take a close look at $l(z, w, -1)$.
From Eq.~(\ref{eq:lztheta_dl}), it is easy to check that
$l(z, w, -1)$ approaches $0$ when $z$ approaches $0$ and,
as long as $w < 0$,
$l(z, w, -1)$ also approaches $0$ when $z$ approaches infinity.
So $l(z, w, -1)$ has a maximum or minimum at some redshift where its
differential with respect to $z$ is equal to $0$.
See Fig.~\ref{fig:lw_sc} (left panel) for examples of $l(z, w, -1)$
versus $z$ for some values of $w$.
For a given $w = w_{\mathrm{np}}$, we can find the redshift, say
$z_{\mathrm{data}}$, where
$\mathrm{d} l(z, w, -1) / \mathrm{d} z = 0$.
Since $\sigma_l^2$ is the variance of $l(z, w, -1)$ along the
redshift, if we only use standard candles distributing in a narrow
redshift range around $z_{\mathrm{data}}$, then $\sigma_l$, as a
function of $w$, will show a local minimum at $w = w_{\mathrm{np}}$
in addition to at the fiducial value $w = -1$.
So, from Eq.~(\ref{eq:Ltheta}), we know that the marginal likelihood
of $w$, $\mathcal{L}(w)$, will have local maxima
at both $w = w_{\mathrm{np}}$ and $w = -1$.
See Fig.~\ref{fig:sl_P_w} for an illustration of two examples.
The fiducial value $w = -1$ is what we want from the constraining,
while $w = w_{\mathrm{np}}$ is an irrelative non-physical value.
A blind analysis without considering the impact of $w_{\mathrm{np}}$
could lead to biased conclusions in the constraining of the dark
energy EOS.
The relation between
$w_{\mathrm{np}}$ and $z_{\mathrm{data}}$
for standard candles is plotted in Fig.~\ref{fig:lw_sc}
(right panel).
We can see that $w_{\mathrm{np}}$ increases with the increase of
$z_{\mathrm{data}}$ and crosses the fiducial value $w = -1$
at $z_{\mathrm{data}} \simeq 1.3$.
This means that, taking into account both the likelihood peaks
corresponding to the fiducial value $-1$ and the non-physical value
$w_{\mathrm{np}}$, as was done implicitly in related analyses,
the standard candles with redshifts less/greater than $1.3$
would appear to favor a dark energy EOS less/greater than $-1$.
This is illustrated in Fig.~\ref{fig:sl_P_w}.

\begin{figure*}[htbp]
  \centering
  \includegraphics[width = 0.24 \textwidth]{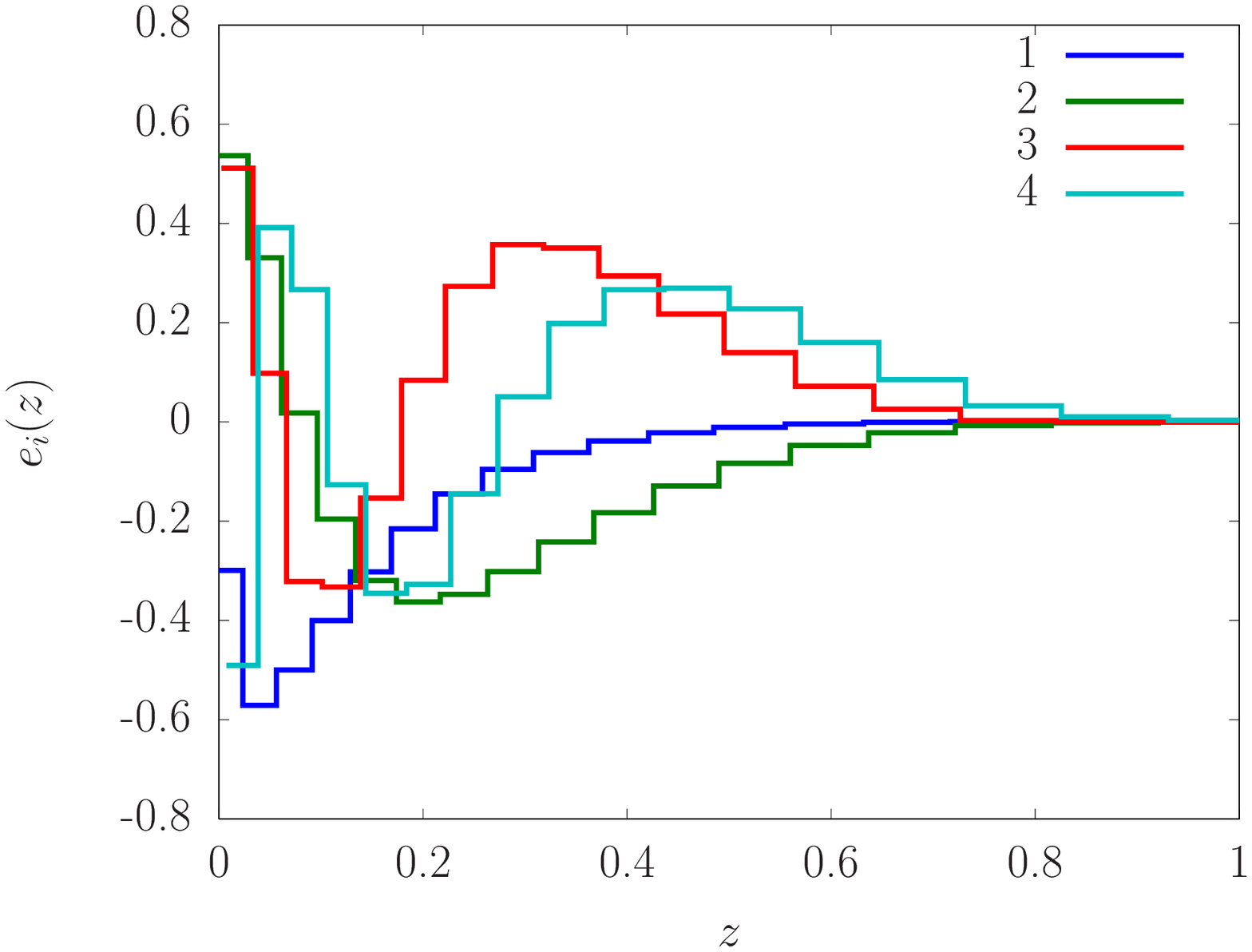}
  \includegraphics[width = 0.24 \textwidth]{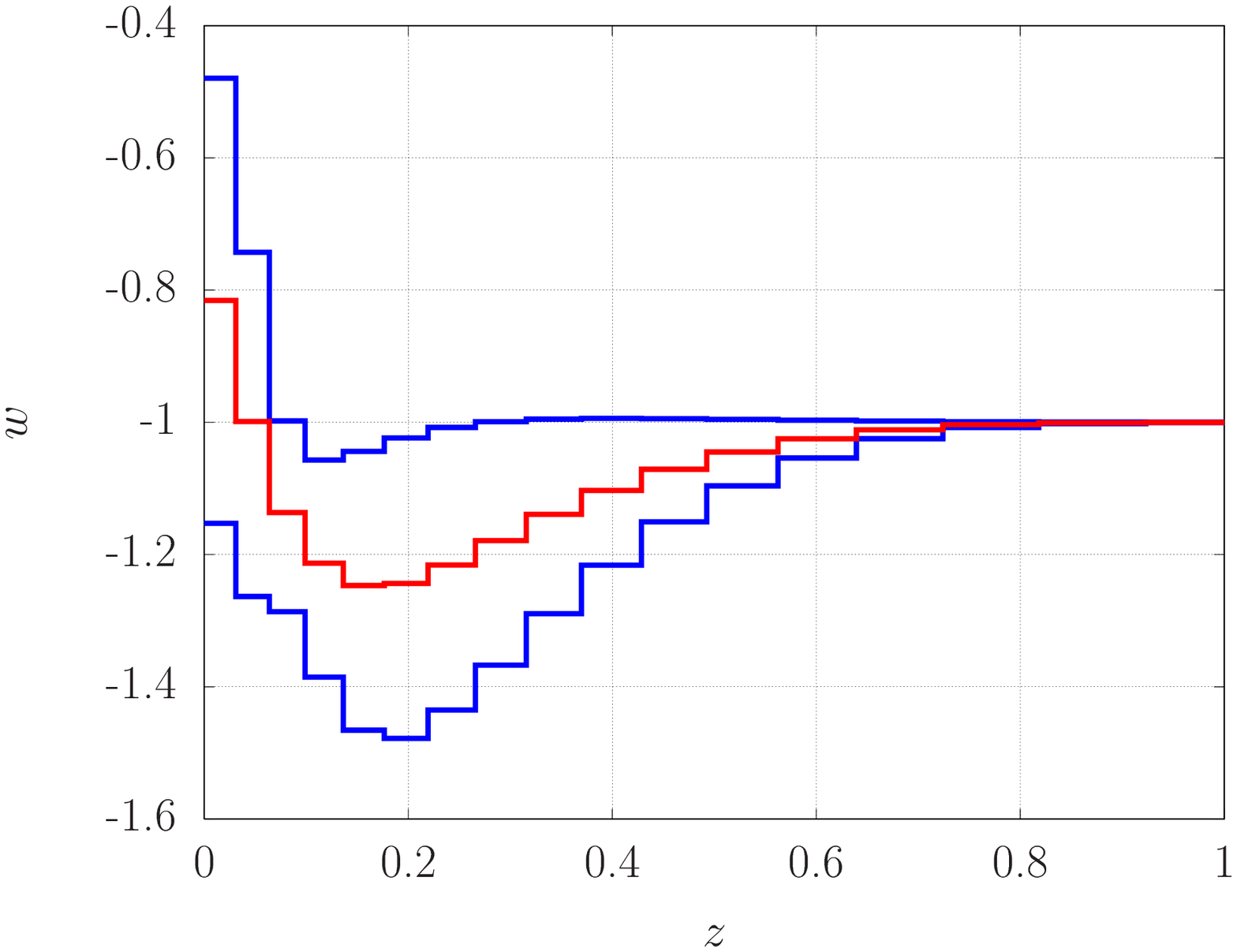}
  \includegraphics[width = 0.24 \textwidth]{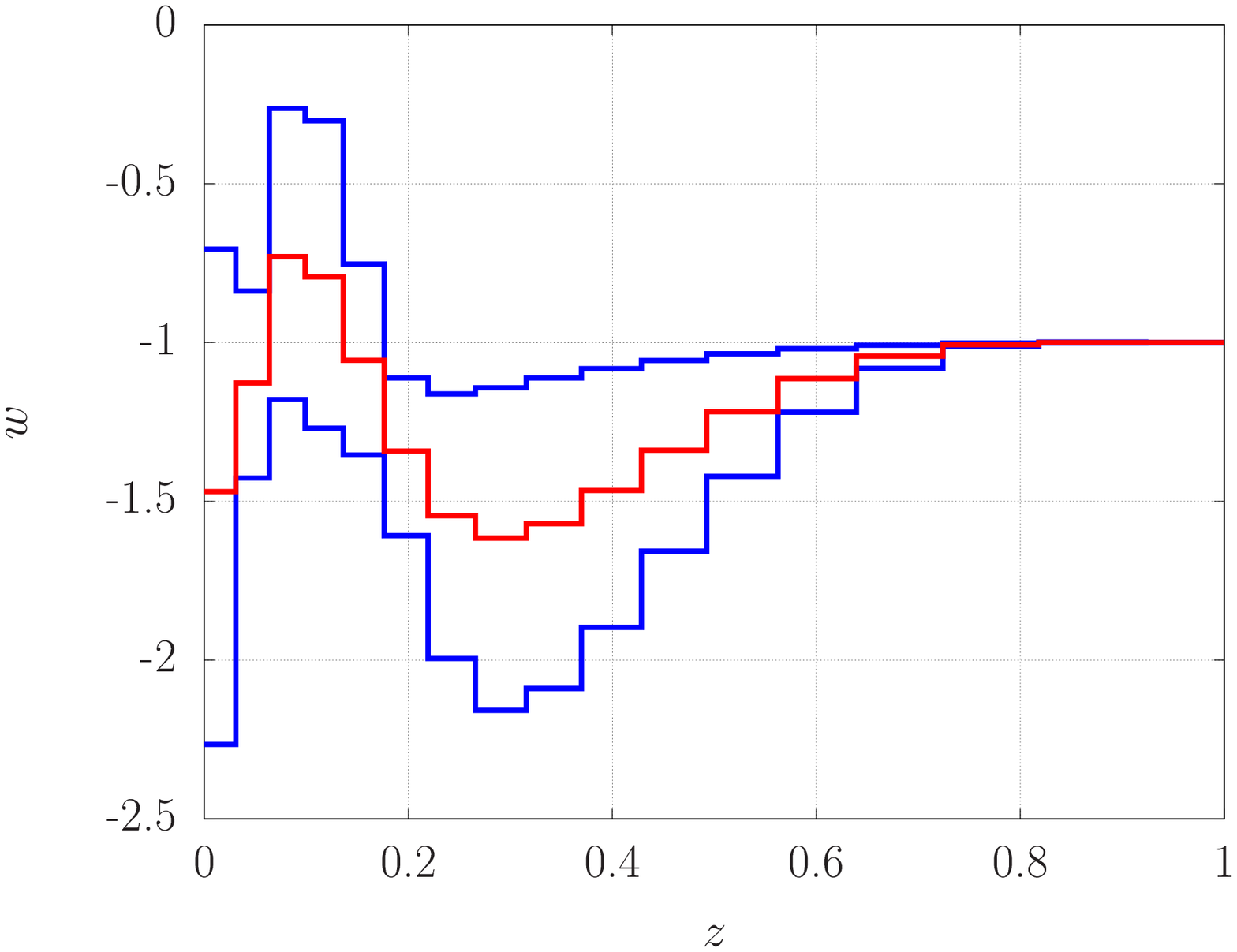}
  \includegraphics[width = 0.24 \textwidth]{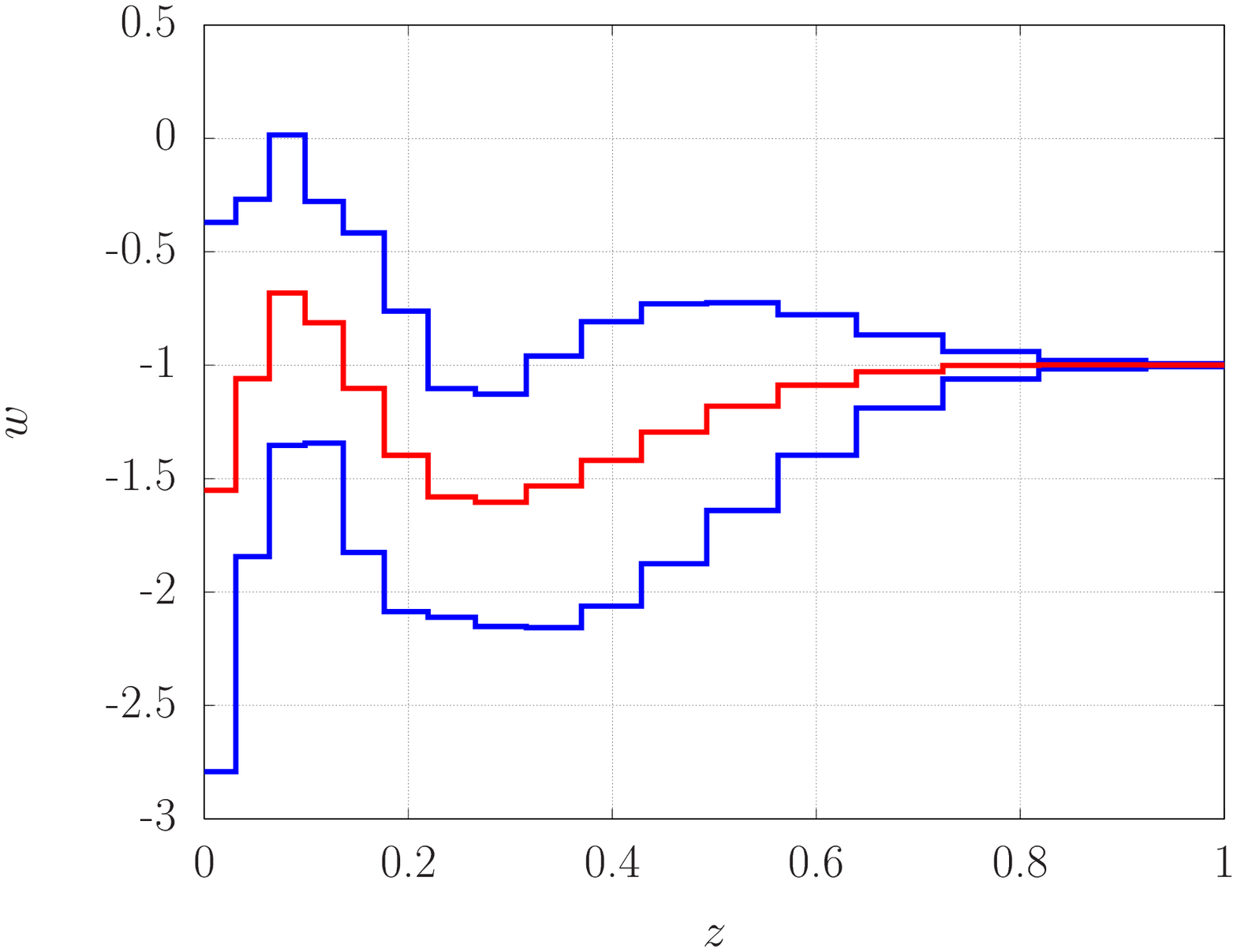}
  \caption
  {
    Principal component analysis~\cite{Huterer:2002hy} of $1+w(z)$.
    $w(z)$ was parameterized using $20$ bins with a constant bin
    width of $0.03$ in scale factor within $[0.4, 1]$
    and $\vert 1+w(z) \vert \leq 50$ was assumed.
    Mock SN Ia data
    ($\sigma_{\mathrm{int}, 0} = 0.06$ and $p = 4$)
    together with mock GRB data
    ($\sigma_{\mathrm{int}, 0} = 0.4$ and $p = 3$)
    were used.
    The mock SN Ia data has the same redshifts of SNLS3
    data~\cite{Conley:2011ku} and
    the mock GRB data has the same redshifts of the sample
    in~\cite{Wang:2011bx}.
    From left to right, the $4$ best coinstrained components of
    $1+w(z)$ are plotted in the first panel
    (Small shifts in $z$ are applied to the components to reduce the
    overlap of the lines so that the components are presented more
    clearly).
    In the following panels, the median values (red lines) and the
    $1\sigma$ confidence intervals (blue lines) of $w(z)$
    reconstructed from the $2$, $3$, and $4$ best constrained
    components of $1+w(z)$ are plotted.
  }
  \label{fig:pca}
\end{figure*}

With that the likelihood peak corresponding to $w_{\mathrm{np}}$
evolves with $z_{\mathrm{data}}$ while the one corresponding to the
fiducial value does not,
the impact of $w_{\mathrm{np}}$ actually would not be an issue if we
use a sample of standard candles with a wide redshift distribution and
assume a constant dark energy EOS,
since the likelihood peak corresponding to the fiducial value would be
strengthened by standard candles at different redshifts while the one
corresponding to $w_{\mathrm{np}}$ would be suppressed.
However, when a flexible parameterization of the dark energy EOS
allowing evolution along the redshift is used,
which is a realistic demand to reconstruct the dark energy EOS
from observation,
the impact of $w_{\mathrm{np}}$ enters since, for these cases, the
constraint of the dark energy EOS at a certain redshift mainly comes
from only part of the standard candles at some redshifts.
For $z_{\mathrm{data}}$ near to $0$,
$w_{\mathrm{np}}$ is far below the fiducial value $w = -1$,
its impact in constraining usually has been implicitly eliminated by
assuming that the dark energy EOS is not very far away from $-1$.
Thus, if the impact of $w_{\mathrm{np}}$ is significant enough,
we can expect, from the constraining, a dark energy EOS that is
around $-1$ at low redshifts, slightly biased to below $-1$ at medium
redshifts, and slightly biased to greater than $-1$ at high
redshifts.
This is exactly the trend that is observed in different
analyses~\cite{Zhao:2012aw, Qi:2008zk, Qi:2009yr, Wang:2011bx,
  Hu:2014ega, Zheng:2014ara, Sullivan:2007tx}.
The significance of the biased trend depends on,
in addition to the redshift distribution of the data,
the flexibility of the parameterization of the dark energy EOS.
A less flexible parameterization means a stronger prior introduced,
which may erase the bias if it is strong enough
(an extreme example is the parameterization of a constant dark energy
EOS along the redshift).
However, a too flexible parameterization usually means large
statistical errors,
which could totally overwhelm the bias and make it not notable.
Especially for cases with a poorly constrained dark energy EOS at high
redshifts,
the likelihood there usually has a long wing at the lower end
(since the data does not care much about the value of the dark energy
EOS at high redshifts as long as the matter dominance is guaranteed,
which means only a upper limit is imposed on the dark energy EOS),
which could make the dark energy EOS at high redshifts appear
less than $-1$ instead of greater than $-1$.
The relation between the significance of the trend and the
parameterization of the dark energy EOS is very subtle.
A principal component analysis~\cite{Huterer:2002hy} of $1+w(z)$
utilizing Eqs.~(\ref{eq:lztheta_dl}) and (\ref{eq:Ltheta})
is presented in Fig.~\ref{fig:pca}.
It shows that the best constrained components do reflect the impact of
$w_{\mathrm{np}}$ with $w$ biased toward less than $-1$ at medium
redshifts, which is consistent with the result of similar analyses
derived from real data~\cite{Zheng:2014ara}.
(The best constrained components do not include the variation of the
dark energy EOS at high redshifts, so the bias from $w_{\mathrm{np}}$
at high redshifts does not show up here.)
It can be expected that,
if the prior introduced by some parameterization of the dark energy
EOS and/or extra data effectively suppress the poorly constrained
components while retaining the best constrained components,
the bias from $w_{\mathrm{np}}$ will be visible in the final result.
For example, Fig.~2 (Panels A1, A2) in~\cite{Zhao:2012aw} shows that
only the 2(3) strongest data modes survive the addition of the prior
there,
thus, when compared with the result here derived from the fiducial
model of the cosmological constant,
the favor of the dynamical dark energy in~\cite{Zhao:2012aw} is
understandable as a result of the impact of $w_{\mathrm{np}}$.

It should be also noted that,
when using Eqs.~(\ref{eq:lztheta_dl}) and (\ref{eq:Ltheta}),
the problem has been simplified in a few places.
For example, the measurement uncertainties are ignored and the
covariance between $x$, $\varepsilon$, and $l$ are assumed to be
exactly zero (Though it is expected to be true, real data are usually
not so ideal).
How these simplifications impact on the constraining of the evolution
of the dark energy EOS need further investigation.
For a precise comparison with results derived from real data, they may
have an unignorable effect and should be taken into account.
And in practice, standard candles are usually handled differently in
different analyses before used to constrain cosmological parameters,
which is another thing need to be considered.

\section{Generalization}

It is easy to see that the above discussions can be applied to more
than just luminosity distance and standard candles.
In fact, Eq.~(\ref{eq:y_dl}) can be generalized to
\begin{equation}
  \label{eq:y_X}
  y = \log
  \left[
    X(z, \theta) \mathcal{F}
  \right]
  ,
\end{equation}
where $X(z, \theta)$ can be anything to be measured from the
observation that depends on cosmological parameters.
All the subsequent derivations are of the same,
except that the definition of $l(z, \theta, \theta_0)$,
Eq.~(\ref{eq:lztheta_dl}), is generalized to
\begin{equation}
  \label{eq:lztheta_X}
  l(z, \theta, \theta_0)
  =
  \log
  \frac
  {
    X(z, \theta)
  }
  {
    X(z, \theta_0)
  }
  .
\end{equation}
The form of Eq.~(\ref{eq:Ltheta}) remains unchanged.
Thus, the results are generalized to other observations.

\begin{figure}[htbp]
  \centering
  \includegraphics[width = 0.23 \textwidth]{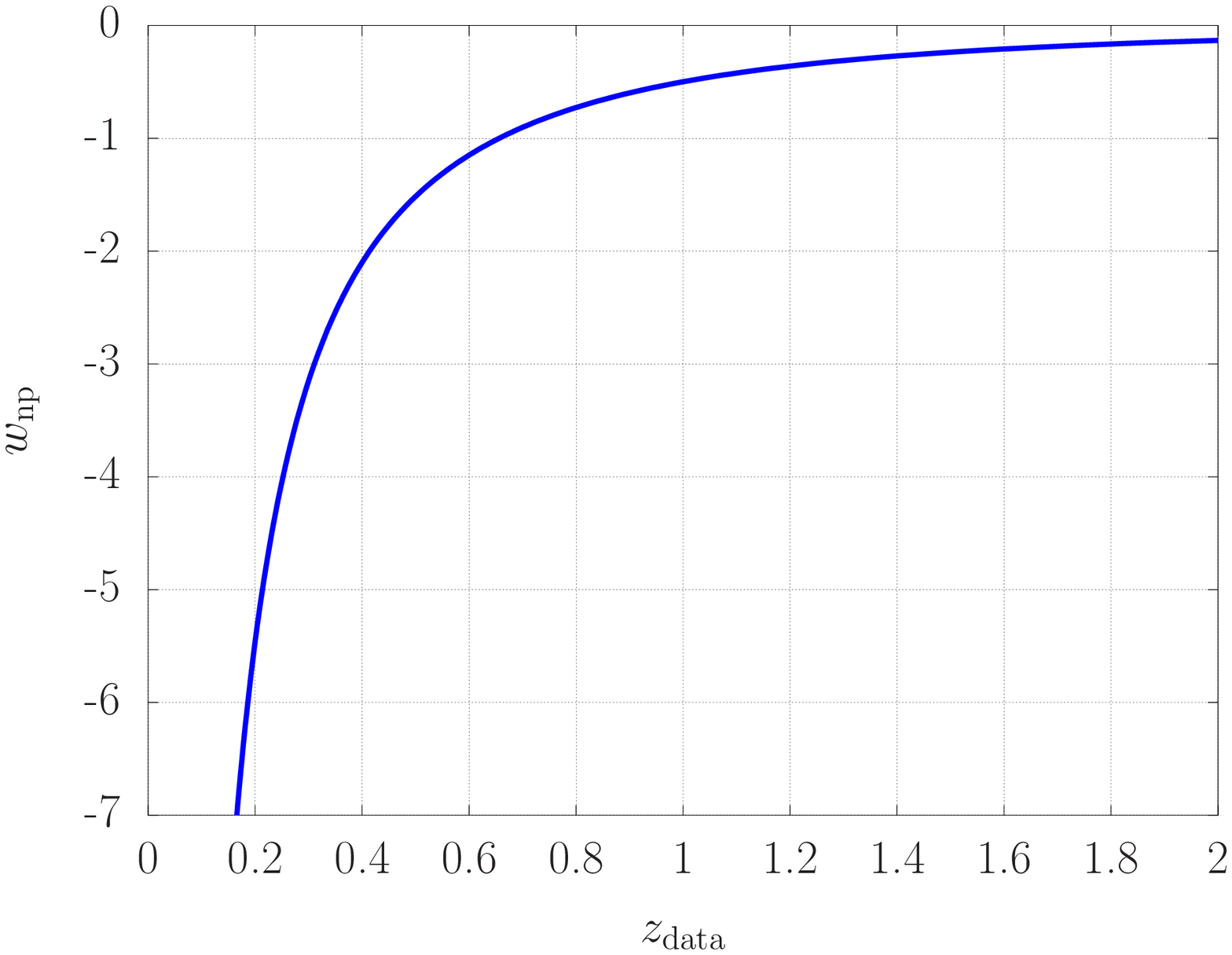}
  \includegraphics[width = 0.23 \textwidth]{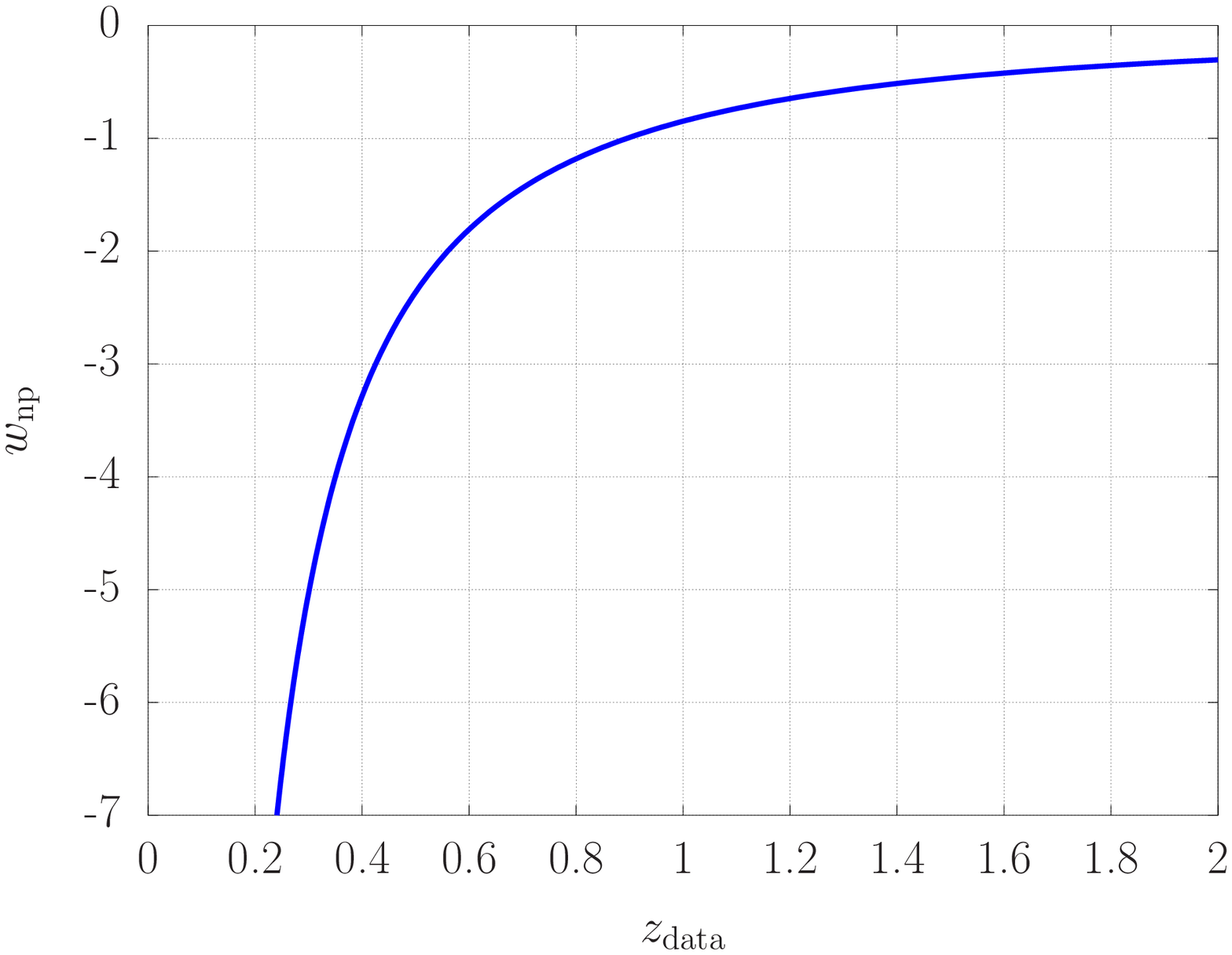}
  \caption
  {
    $w_{\mathrm{np}}$ versus $z_{\mathrm{data}}$
    for the standard ruler of the line-of-sight direction (left)
    and of the combined directions (right).
  }
  \label{fig:lw_bao}
\end{figure}

For BAO survey, the comoving sound horizon at the baryon drag epoch,
$r_s(z_{\mathrm{drag}})$, is used as the standard ruler.
For the standard ruler of the transverse direction, of the
line-of-sight direction, and of the combined directions, the following
distance ratios are measured from the survey respectively:
\begin{IEEEeqnarray}{rCl}
  \theta_s
  &=&
  \frac
  {
    r_s(z_{\mathrm{drag}})
  }
  {
    (1+z) d_A
  }
  ,
  \\
  \delta z_s
  &=&
  \frac
  {
    r_s(z_{\mathrm{drag}}) H(z)
  }
  {
    c
  }
  ,
  \\
  d_s
  &=&
  \frac
  {
    r_s(z_{\mathrm{drag}})
  }
  {
    D_V
  }
  .
\end{IEEEeqnarray}
These relations can be rewritten in the form of $y = a + \varepsilon$
with $a = \log [r_s(z_{\mathrm{drag}})]$,
$\mathcal{F}$ being $\theta_s$, $\delta z_s$, and $d_s$ respectively,
and $X(z, \theta)$ being $(1+z) d_A$, $c / H(z)$, and $D_V$
respectively.
Thus, $l(z, \theta, \theta_0)$ can be calculated from
Eq.~(\ref{eq:lztheta_X}) and corresponding $w_{\mathrm{np}}$ can be
derived from it.
(\emph{Here, only BAO measurements from density fluctuations of
  baryonic matter are considered, no prior information about
  $r_s(z_{\mathrm{drag}})$ from cosmic microwave background (CMB)
  measurements is inputted.
  $r_s(z_{\mathrm{drag}})$ is treated as an unknown constant and
  simply marginalized during the constraining.
  For example, in~\cite{Percival:2007yw, Percival:2009xn}, part of the
  BAO measurements are summarized in distance ratio
  $D_V(0.35)/D_V(0.2)$, which is effectively equivalent to
  marginalizing out $r_s(z_{\mathrm{drag}})$ if used to constrain the
  dark energy EOS.
  If BAO measurements are combined with CMB measurements, such that
  the information about $r_s(z_{\mathrm{drag}})$ is inputted, the
  impact of $w_{\mathrm{np}}$ should be relieved or eliminated.})
Since the luminosity distance $d_L$
and the angular diameter distance $d_A$
relate to each other though $d_L = d_A (1+z)^2$,
the $l(z, \theta, \theta_0)$ for the standard ruler of the transverse
direction differs from that for standard candles
only by a constant factor of $2$.
So the derived $w_{\mathrm{np}}$ is totally the same.
For the other two cases, i.e., the standard ruler of the line-of-sight
direction and of the combined directions,
% $l(z, w, -1)$ versus $z$ for some values of $w$ and
$w_{\mathrm{np}}$ versus $z_{\mathrm{data}}$
is plotted in Fig.~\ref{fig:lw_bao}.
We can see similar behaviors of $w_{\mathrm{np}}$ versus
$z_{\mathrm{data}}$.
$w_{\mathrm{np}}$ crosses the fiducial dark energy EOS at
different values of $z_{\mathrm{data}}$.
For the former case, at $z_{\mathrm{data}} \simeq 0.65$,
and for the latter one, at $z_{\mathrm{data}} \simeq 0.9$.

To be more general, what were discussed here is not limited to dark
energy or cosmology.
In fact, it represents a class of mathematical problems
of Bayesian analysis.
Fitting data to a linear relation and using the relation to constrain
model parameters are very common in data analyses.
The key quantity in the discussions, $\sigma_l$, appears during
the marginalization (it shows up in~\ref{enum:intercept} through
$\sigma_m$ when we integrate over the intercept parameter),
which is almost inevitable in data analyses.
So similar biases may show their shadow here and there in
similar problems.
Concerning data analyses in cosmology, a lot of parameters are
involved.
Marginalization is used very often.
The kind of bias discussed here should be checked carefully in the era
of precision cosmology.

\section{Summary}

Starting with luminosity relations of standard candles, the steps was
described that one should follow to estimate the constraints of a
luminosity relation on cosmological parameters using mock data,
from which a simple and general formula was deduced that can be used
to directly calculate the marginal likelihood of cosmological
parameters.
Using the formula together with the dependence of the cosmic expansion
on the dark energy EOS, it was shown how a kind of bias could arise
that leads to a fake evolution of the dark energy EOS,
whose significance depends on the flexibility of the parameterization
of the dark energy EOS and the redshift distribution of the data.
Then the formula was generalized to more than just standard candles.
It was shown that the BAO data could lead to similar biases. It was
mentioned that the issue represents a class of mathematical problems
of Bayesian analysis and should be paid attention to in similar
analyses.

\begin{acknowledgments}
  This work was supported by the National Natural Science Foundation
  of China (Grant No.~10973039, 11203079, and 11373068),
  Key Laboratory of Dark Matter and Space Astronomy (Grant
  No.~DMS2011KT001), and the Project of Knowledge Innovation Program
  (PKIP) of Chinese Academy of Sciences (Grant No.~KJCX2.YW.W10).
\end{acknowledgments}

\bibliographystyle{apsrev4-1}
\bibliography{dark_energy,grb,misc}

\end{document}